\documentclass[aps,prapplied,twocolumn,floats,showpacs,floatfix,superscriptaddress]{revtex4-1}
\usepackage{graphics,graphicx,dcolumn,bm,float}
\usepackage{amssymb,amsmath,multirow,rotate,color,float}
\usepackage[latin1]{inputenc}
\usepackage{color}
\usepackage{epstopdf}

\usepackage[dvips]{epsfig}

\begin{document}

\title{Superfast collective motion of magnetic particles}

\author{Olga Baun}
\affiliation{Institut f\"ur Physik, Johannes Gutenberg-Universit\"at
  Mainz, 55099 Mainz, Germany}

\author{Peter Bl\"umler}
\email[Corresponding author: ]{bluemler@uni-mainz.de}
\affiliation{Institut f\"ur Physik, Johannes Gutenberg-Universit\"at
  Mainz, 55099 Mainz, Germany}

  \author{Friederike Schmid}
\affiliation{Institut f\"ur Physik, Johannes Gutenberg-Universit\"at
  Mainz, 55099 Mainz, Germany}

   \author{Evgeny S. Asmolov}
\affiliation{A.N.~Frumkin Institute of Physical
Chemistry and Electrochemistry, Russian Academy of Sciences, 31
Leninsky Prospect, 119071 Moscow, Russia}
\affiliation{Institute of Mechanics, M.V.~Lomonosov Moscow State
University, 119991 Moscow, Russia}

\author{Olga I. Vinogradova}
\email[Corresponding author: ]{oivinograd@yahoo.com}
\affiliation{A.N.~Frumkin Institute of Physical
Chemistry and Electrochemistry, Russian Academy of Sciences, 31
Leninsky Prospect, 119071 Moscow, Russia}
\affiliation{Department of Physics, M.V.~Lomonosov Moscow State University, 119991 Moscow, Russia}
\affiliation{DWI - Leibniz Institute for Interactive Materials, Forckenbeckstra\ss e 50, 52056 Aachen,
  Germany}

\date{\today}
\begin{abstract}

It is well-known that magnetic forces can induce a formation of densely packed strings of magnetic particles or even sheafs of several strings (spindles). Here we show that in a sufficiently strong magnetic field, more complex aggregates of
 particles, translating with a much faster speed than would be for a single particle or even a spindle, can be assembled at the water-air interface. Such a superfast flotilla is composed of many distant strings or spindles, playing a role of its vessels, and moves, practically, as a whole. We provide theoretical results to interpret the effect of a collective motion of such magnetic vessels. Our theory shows that, in contrast to an isolated chain or spindle, which velocity grows logarithmically with the number of magnetic particles, the speed of the interface flotilla becomes much higher, being proportional to the square root of their number. These results may guide the design of magnetic systems for extremely fast controlled delivery.

\end{abstract}

\maketitle

\section{Introduction}

\begin{figure}[tb]
\includegraphics[width=1\columnwidth]{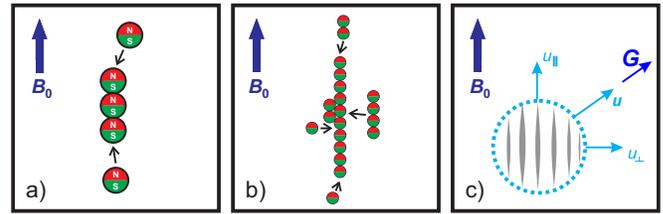}
  \caption{Schematic presentation of particle aggregation: a) particles orient
along the external homogeneous magnetic field, $\mathbf{B}$, and form strings by mutual
dipolar attraction. b) Once a certain length is
reached \protect\cite{heinrich.d:2011}, particles or smaller particle chains can
also attach towards the center of the string to form spindles. c) Strings and spindles
 can form larger assemblies (indicated by the circle and referred to as flotilla). The
flotilla moves as a whole when a field gradient $\mathbf{G}$ is applied with a velocity
$u$. Due to the fact that the chains always align parallel to  $\mathbf{B}_0$
the velocity is anisotropic with components $u_{\parallel }$ and $u_{\perp }$.}
\label{fig:intro1}
\end{figure}

Magnetic micro- and nanoparticles have received much attention in recent years.
They are important for a variety of promising applications including data
storage, targeted drug delivery and cancer diagnostics/treatment
systems~\cite{singamaneni.s:2011, tierno.p:2014}. Most recent work has focused
on the synthesis and biomedical modifications of the
particles~\cite{singamaneni.s:2011} or their detection and separation~\cite{zborowski1999,pamme2006}. Their field-directed assembly from
dispersions to aggregates has also been studied for many
years~\cite{martin.je:1999, furst.em:2000, lalatonne.y:2004, taheri.sm:2015,
li.yh:2016, klapp.shl:2016}. In typical applications, a magnetic field is
applied, which does not change dramatically on the length scale of the
aggregates. Hence, the magnetic moments of neighboring particles point in the
same direction and exert dipolar forces upon each other, causing the formation
of strings along the flux lines (see  Fig.~\ref{fig:intro1} (a)). This is
analogous to the behavior of iron filings in classical demonstration
experiments. If such strings are allowed to grow further until they reach a
critical average length~\cite{heinrich.d:2011, heinrich.d:2015}, they form
spindle-like aggregates (see Fig.~\ref{fig:intro1} (b)), which at a certain size
exert a repelling force on the neighboring string or spindle again, analogous to the
familiar picture of iron filings in magnetic fields.
Apart from inducing aggregation, time-dependent (e.g., rotating) magnetic
fields and magnetic field gradients can also be used to manipulate and drive isolated particles~\cite{wu.l:2009, zhou.r:2017} and their aggregates~\cite{melle.s:2003, belkin2007, martin.je:2013, straube2014, martinez2015}.

The body of work investigating motion of particles in magnetic fields still remains rather scarce, although there has been some interest in recent years in questions surrounding the dynamic behavior of magnetic particles in microfluidic channels and at the liquid-air or solid-liquid interface. Questions of interest included the translation of isolated particles~\cite{shevkoplyas2007}, the controlled motion of chains at the periodically rough solid~\cite{straube2014}, the propulsion of colloidal microworms~\cite{martinez2015}, the formation of dynamic patterns  on a fluid surface~\cite{belkin2007}. Most of the studies calculate the translational velocity by balancing the
magnetic force on a single particle and the Stokes drag~\cite{martin.je:1999, melle.s:2003, lalatonne.y:2004, heinrich.d:2011,
heinrich.d:2015}. In other words, it is normally assumed that the motion of a particle is unaffected by the presence of other particles. However, recent experiment~\cite{baun2017} has demonstrated that  magnetic particles suspended at the fluid interface can move much faster than one would expect for an isolated particle, but these results have not been interpreted, and the understanding of such a superfast motion is still challenging. Clearly, in order to rationalize the situation, new experimental data and fresh theoretical concepts are necessary.


In this paper, we explore experimentally the fast motion of superpara/magnetic particles suspended at the water-air interface and suggest simple theoretical models describing the data. Our results  show that strings or their clusters (spindles) form, but then assemble in a more complex aggregate referred to as a flotilla, where many distant chains or spindles, playing a role of vessels, move together practically as a whole (see Fig.~\ref{fig:intro1}(c)). The velocity of such a flotilla is orders of magnitude higher than that of a single particle, and is even much faster than for an isolated vessel. Our simple models describe well the velocities of isolated vessels and flotillas, relating them to the size of the aggregates, the parameters of constituting particles and their number.  The formation of a fast flotilla of magnetic particles could greatly reduce the time required for controlled particle manipulations. This may have important consequences for controlled delivery of magnetic particles in various biomedical and other applications.


\section{Experimental}
\label{sec:experimental}

\begin{figure}[tb]
\includegraphics[width=0.9\columnwidth]{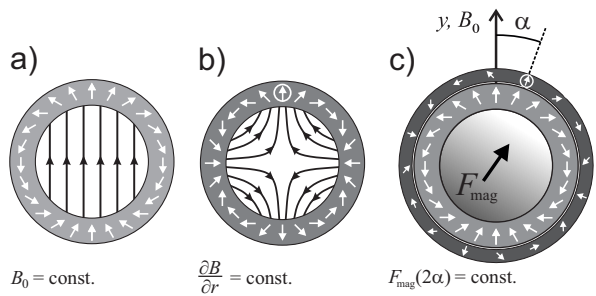}
  \caption{Schematic drawing of a cross-section through a) an ideal Halbach dipole cylinder providing a homogeneous magnetic field, $\mathbf{B}_0$;
  b) an ideal Halbach quadrupole generating a constant radial magnetic field gradient. c) If the two are superimposed, only
  the gradient component along $\mathbf{B}_0$ is relevant to generate a magnetic force, $\mathbf{F}$, on particles
  in the inner volume. The angle of $\mathbf{F}$ changes by $2\alpha $ if the quadrupole is rotated by $\alpha $ relative to the dipole. }
  \label{fig:exp1}
\end{figure}

The experiments were carried out using the same technique as in a recent study by two of us~\cite{baun2017}.   Its basic idea is to use a magnet
system which provides a strong, homogeneous, dipolar magnetic field
($\mathbf{B}_0$) (cf.  Fig.~\ref{fig:exp1}(a)) to magnetize and orient the superparamagnetic particles (SPPs)
(as shown in Fig.~\ref{fig:intro1}), and a second quadrupolar field with
a spatially constant tensor $\nabla \mathbf{B}$,
superimposed on the first, to generate a magnetic force on the oriented
particles (see Fig.~\ref{fig:exp1}(b)). In this configuration the motion of the
particles is driven predominantly by the gradient $\mathbf{G}$ of the field
component in the direction of the homogeneous field, $\mathbf{G} = \nabla
(\mathbf{B} \cdot \mathbf{B}_0)/B_0$.  As a result, particles are guided with
constant force and in a single direction over the entire inner volume of the
magnet. The direction of the force is simply adjusted by varying the angle
between the quadrupole and the dipole (see Fig.~\ref{fig:exp1}(c)). Since a
single magnetic field gradient is forbidden by Gauss' law (the tensor $\nabla
\mathbf{B}$ must be traceless), the other gradient component of the quadrupole
determines the angular deviation of the force, which is negligible for
$|\mathbf{B}_0|\gg |\mathbf{G}|r$. A possible realization of this idea is a coaxial
arrangement of two Halbach cylinders \cite{halbach1980,blumler2015}. A Halbach
dipole to evenly magnetize and orient the particles, and a Halbach quadrupole
to generate the magnetic force, $\mathbf{F}$, on the SPPs.

 \begin{figure}[tb]
\includegraphics[width=0.4\columnwidth]{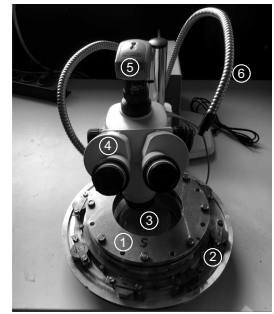}
  \caption{Photograph of the experimental device: 1) Halbach dipole, 2) Halbach quadrupole (can be rotated), 3) Opening with Petri-dish, 4) stereo-microscope, 5) CCD-camera, 6) incident lightning on each side}
  \label{fig:setup}
\end{figure}

 The magnet system used is shown in Fig.~\ref{fig:setup}. Design and characterization
of this magnet system is described in detail in \cite{baun2017}. This system has an
inner opening of 130 mm in diameter in which a Petri-dish (70 mm in diameter
and 20 mm height) was centered. All experiments were done in the central area
of this dish (ca. 40 mm in diameter), where the magnetic field has the
following properties: $\textit{B}_0$ = 0.103(2) T and $\textit{G}$ = 0.19(2)
T/m. The Petri-dish was filled with ca. 30 ml of deionized water ($\mu$ = 0.9
mPs).

To understand how the speed of the aggregate depends on the total number of constituent magnetic particles, which is determined optically, particles must be micron-sized. Here we use a
commercial
(micromod Partikeltechnologie GmbH, Rostock, Germany) sample of spherical
magnetite (ca. 40\% w/w) particles in a matrix of poly(D,L-lactic acid) with a
molecular weight of 17 kD and a plain surface (product name:  PLA-M 30 $\mu$m and
product number: 12-00-304) with a size distribution range: 20-50 $\mu$m, centered at
30 $\mu$m; magnetization $M$ = 4.3 Am$^2$/kg particles for $B =$ 0.1 T;
saturation magnetization $M >$ 6.6 Am$^2$/kg particles for $B >$ 1 T;
density: 1300 kg/m$^3$, delivered as an aqueous suspension with  $5.5 \times
10^5$  particles/ml).


 From the stirred suspension 10 $\mu$l were extracted and injected into the
Petri-dish using an Eppendorf pipette (Wetzlar, Germany).  However, not all of
the (theoretically expected) 5500 particles in this volume could be transferred
into the water inside the dish because several stuck to the pipette,
additionally some directly sunk to the ground. Smaller concentrations or
volumes were found to be unpractical because too many particles stuck at the
pipette tip. In order to obtain aggregates formed by a certain number (order
of magnitude) of particles, some of the strings were removed again from the
sample by sucking them into another clean pipette. We note that the particles
and particle aggregates studied here (those that did not sink to the ground)
are confined at the air/water interface, hence the systems are effectively
two dimensional.

 The composition of the aggregates was studied using a stereo microscope (Motic
SMZ-168, Wetzlar, Germany) with a modest magnification (15 $\times$ to
100$\times$) and incident illumination. The third tube of the microscope was equipped with
a CCD-camera (moticam 1000, motic, Wetzlar, Germany) allowing to take snapshots
as well as real-time videos. Once an aggregate of suitable size was formed and
singled out, the number of particles, $\textit{n}$, it is composed of was
determined via the microscope. For smaller numbers ($\textit{n}\lesssim$ 300)
this could be done at relative high magnifications where single particles could
be identified.  The composition of larger structures had to be estimated from
graphical integrals of their shapes, which of course increased the error for
larger $\textit{n}$.  The selected aggregate is then moved parallel and
perpendicular to $\textit{B}_0$ by rotating the quadrupole in steps of
45$^{\circ}$ (hence the forceand the particle movement happens in steps of
90$^{\circ}$). In the lowest magnification the microscope has a field of
view of 4 $\times$ 4 mm$^{2}$ which is too small to study the typical
velocities of the aggregates (except for very small $\textit{n}$). Therefore, a
digital camera (Fujifilm FinePix HS30EXR) was used to take real-time movies of
this motion.  A transparent foil with a millimeter-grid was attached to the
bottom of the Petri-dish as a spatial reference.  Particle and aggregate
velocities were obtained from these videos using the image processing software
Avidemux V5 (NCH Software, Greenwood Village CO, USA).

\section{Results and discussion}

\begin{figure}[b]
\includegraphics[width=0.9\columnwidth]{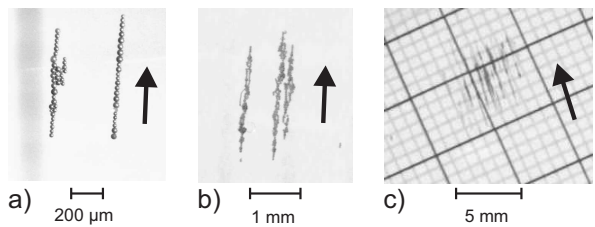}
  \caption{Snapshots of typical aggregates from different amounts of 30 $\mu$m magnetite particles: a) typical  vessel formation ($n \lesssim$ 100). The assembly on the right is a string, while on left
  side already some particles attach close to the center (spindle). b) spindle-shaped vessels for larger $n$ and c) a "flotilla" of such vessels.
  The black arrows indicate the direction of the homogeneous magnetic field $\mathbf{B}_0$. }
  \label{fig:exp2}
\end{figure}

Shortly after the injection of the particles, strings and later spindles form
from the initial cloud of particles (see Fig.\ \ref{fig:exp2}). The aggregation occurs quasi instantly,
but note that for nanoparticles it could take several
minutes. After that  several parallel distant vessels form a flotilla.  As explained
in the introduction, the reorganization of the initially dispersed particles into
vessels (strings or spindles)  is driven by magnetic dipolar
interactions. The mechanism that keeps the flotilla together is less clear.
Whereas the magnetic long-range interactions between parallel rigid strings
made of identical dipolar particles are repulsive, fluctuations in flexible
chains may induce long-range attractive interactions {\em via} a mechanism that
was first pointed out by Halsey and Toor \cite{halsey.tc:1990}. However, this
mechanism would typically induce chain coalescence \cite{furst.em:2000} and not
stabilize larger arrays, i.e.  flotillas, of well-separated vessels. Other
possible sources of attraction are chain defects \cite{martin.je:1999} or
long-range capillary interactions mediated by the water-air interface
\cite{dominguez.a:2007, bresme.f:2007}.

The aggregates can be set in motion by applying an additional (quadrupolar)
field with constant gradient $G$ as explained in Sec.\ \ref{sec:experimental}.
The stationary velocities of the aggregates are found to be orders of magnitude higher
than the velocities calculated for isolated particles. To study this effect more
systematically, we have conducted experiments by translating aggregates
containing different amounts of SPPs parallel and perpendicular to the
direction of $\mathbf{B}_0$.

\begin{figure}[h]
\includegraphics[width=.8\columnwidth]{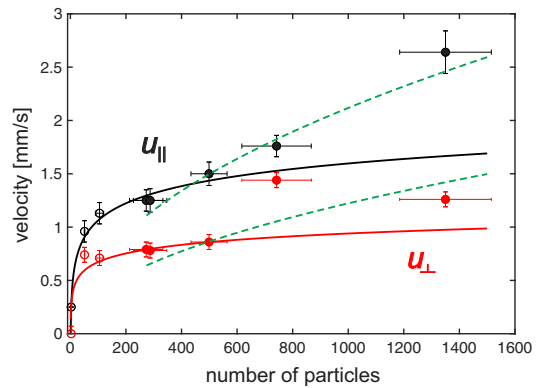}
\caption{
Experimentally obtained longitudinal ($u_{\parallel}$ in black) and transverse
($u_{\perp}$ in red) velocities of vessels and flotillas vs the number of
constituent particles, $n$.
The open circles correspond to single vessels while the solid circles
correspond to flotillas of 2  - 4 strings each (c. f. Table \
\ref{tab:exp}). The solid lines represent fits of Eq.(\ref{eq:cyl_par_2}) (in
black) and Eq.  (\ref{eq:cyl_perp_2})  (in red) to the data, where the radius
of the particles was fitted for both curves as $a = 17.4 \mu$m  $\pm 0.6
\mu$m while the other parameters in Eq. (\ref{eq:C}) were taken from Sec.\
\ref{sec:experimental}.
The dashed green lines show the prediction of Eq.\ (\ref{eq:disk}) for
flotillas only. Here, $\phi$ was adjusted to cover the range of data. The lower
curve corresponds to $\phi = 0.02$ and the upper to $\phi = 0.06$.
See text for more details.
}
\label{fig:results1}
\end{figure}

Table \ \ref{tab:exp}  shows the results of the velocity measurements of various aggregates which are
then plotted in Fig.\ \ref{fig:results1} as a function of the total number of particles (note that here we present data only for relatively small aggregates since it is difficult to count $n$ in larger flotillas, which move even faster). The central observation can be summarized as
follows: Under an applied magnetic force, particles forming a flotilla of
vessels collectively translate at a much higher speed than one would expect for
isolated particles.  Why does this happen and what does this mean?

\newcommand\B{\rule{0pt}{10pt}}
\begin{table}
\caption{Experimentally determined properties of PLA-M SPPs: $n$ is the total
number of particles in the aggregate, $N$ is the number of vessels per flotilla
with $n_{i}$ as the number of particles in its individual vessel. The
velocities parallel,  $u_{\parallel}$, and perpendicular,  $u_{\perp}$,  to
$\mathbf{B}_0$ were determined for each aggregate as described in Sec.  \
\ref{sec:experimental}. The same data are plotted in Fig. \
\ref{fig:results1}.}
\centering{
\begin{tabular} {p{1.8cm}p{2.3cm}p{1.9cm}p{1.9cm}}
\B
$n$  & $n_{i} (i=1..N)$ & $u_{\parallel}$ [mm/s] & $u_{\perp}$ [mm/s] \\
\hline
\B
3 & $n_{1} =3$ & $0.25\! \pm\! 0.01$ & $0\! \pm\! 0.06$\\
$51\! \pm\! 1$ & $n_{1} =51\! \pm\! 1$ &  $0.96\! \pm\! 0.1$ & $0.74\! \pm\! 0.06$\\
$105\! \pm\! 15$ & $n_{1} =105\! \pm\! 15$ & $1.13\! \pm\! 0.1$ & $0.71\! \pm\! 0.06$\\
\B
$273\! \pm\! 60$ & $n_{1} = 12\! \pm\! 25$ & $1.25\! \pm\! 0.1$ & $0.79\! \pm\! 0.07$\\
    &  $n_{2} = 150\! \pm\! 35 $\\
\B
$287\! \pm\! 60$ & $n_{1} = 102\! \pm\! 25$ & $1.25\! \pm\! 0.11$ & $0.78\! \pm\! 0.07$\\
    &  $n_{2} =105\! \pm\! 25 $\\
    &  $n_{3} = 80\! \pm\! 10 $\\
\B
$499\! \pm\! 65$ & $n_{1} = 181\! \pm\! 25$ & $1.5\! \pm\! 0.11$ & $0.86\! \pm\! 0.07$\\
    &  $n_{2} =191\! \pm\! 25 $\\
    &  $n_{3} = 127\! \pm\! 15 $\\
\B
$742\! \pm\! 125$ & $n_{1} = 230\! \pm\! 25$ & $1.76\! \pm\! 0.1$ & $1.44\! \pm\! 0.07$\\
    &  $n_{2} =512\! \pm\! 100 $\\
\B
$1350\! \pm\! 165$ & $n_{1} = 450\! \pm\! 50$ & $2.64\! \pm\! 0.2$ & $1.26\! \pm\! 0.07$\\
    &  $n_{2} =400\! \pm\! 50 $\\
    &  $n_{3} =350\! \pm\! 40 $\\
    &  $n_{4} =150\! \pm\! 25 $\\
\hline
\end{tabular}}
\label{tab:exp}
\end{table}

The detailed transport equations are rather complex, and it is difficult to extract a simple
message from them. Our aim is more modest. We want to highlight the basic
principles of observed phenomena in very compact terms. We model
single isolated vessels by cylinders of radius $a$ and length $2b$, which lie
in the air-water interface and translate under the action of a tangent to this interface external
magnetic force $F_n$. We
consider the hydrodynamic drag caused by the water phase and neglect the
friction with the air phase. Due to the small sizes of the aggregates and their
relatively small velocities (of the order of mm/s), the Reynolds number is small, so that we use Stokes
equations.  We further assume that the air-water interface is strictly planar
and that the floating at the interface particles are half immersed in water.
Physically, this means that we neglect differences of the surface tension of
the particles with water or air and gravitational forces.  With these
assumptions, the solution of the Stokes equation for the hydrodynamic flows
in the half space underneath the aggregates can be obtained from the solution
for a fully immersed aggregate by simply ``cutting away'' the upper half space.
Hence we can use the available results for the resistance of bodies in an
infinitely extended fluid at low Reynolds numbers \cite{happel1965} after
replacing the viscosity of water, $\mu$ by the effective viscosity
$\mu_{\mbox{\tiny eff}} = \mu/2$~\cite{ranger1978}.


In the limit of very long and thin cylinders, $b \gg a$, the translational
velocity in longitudinal and transverse direction (parallel and perpendicular
to the string orientation, i.e. $\mathbf{B}_0$ ), $u_{\parallel}$ and
$u_{\perp}$, can be expressed as (Eqs.\ (5-11.52) and (5-11.54) in Ref.\
\onlinecite{happel1965}),

\begin{eqnarray}
u_{\parallel }& \approx&
  \frac{F_{n}\left[ \ln \left( 2b/a\right) -0.72\right]}
        {4\pi \mu_{\mbox{\tiny eff}} b}.\,
\label{eq:cyl_par} \\
u_{\perp }& \approx &
  \frac{F_{n}\left[ \ln \left( 2b/a\right) +1/2\right] }
        {8\pi \mu_{\mbox{\tiny eff}} b}.\,
\label{eq:cyl_perp}
\end{eqnarray}%

For a single string we naturally set the cylinder radius $a$ as equal to the
particle radius and, therefore, we have $b=na$, where $n$ is the number of
magnetic particles in a single vessel. The force acting on the single vessel
can be calculated as $F_{n}=n|\mathbf{f}|$, where $\mathbf{f}$ is the magnetic
force acting on a single particle. The latter can be expressed as
$\mathbf{f=\nabla (m\cdot B)} $, where the vector $\mathbf{m}$ is the magnetic
moment which is aligned along $\mathbf{B}_0$ in this case,
$\mathbf{m}=\frac{4}{3}\pi a^{3}\rho M(B) \mathbf{B}_{0}/B_{0} $; $\rho$ is the
density and $\textit{M(B)}$ the magnetization (which typically is constant over
space, because $B = B_{0}$ is homogeneous as $G$ only causes a small deviation
on the lengthscale of the aggregates). Hence, the force on a single particle
can be simplified as $\left\vert\mathbf{f}\right\vert =\frac{4}{3}\pi a^{3}\rho
MG.$ By substituting the expression for $F_{n}$ in Eqs.(\ref{eq:cyl_par}) and
(\ref{eq:cyl_perp}) we obtain
\begin{eqnarray}
u_{\parallel }& \approx & 2 C \left[ \ln \left( 2n\right) -0.72\right]
\label{eq:cyl_par_2} \\
u_{\perp }& \approx & C \left[ \ln \left( 2n\right) +0.5\right]
\label{eq:cyl_perp_2} \\
\mbox{with} \qquad
 C &=& a^2 \rho \: M \: G / 6 \mu_{\mbox{\tiny eff}}.
\label{eq:C}
\end{eqnarray}
The important conclusion from Eqs.(\ref{eq:cyl_par_2}) and
(\ref{eq:cyl_perp_2}) is that both $u_\parallel$ and $u_\perp$ grow
logarithmically with $n$.  When the force is applied at some angle $\alpha$ to
the cylinder axis (the  $\mathbf{B}_0$- or $y$-axis) its velocity will be
$\mathbf{u}=u_{\perp }\mathbf{e}_{x}\sin \alpha + u_{\parallel
}\mathbf{e}_{y}\cos \alpha$.
Therefore, the velocity of a vessel (here a string) in any direction scales as $\ln
\left(n\right)$ at large $n$.  This suggests that it should be impossible to obtain ultrafast
velocities just by increasing the number of particles in a single vessel.

Fig.\ \ref{fig:results1} shows a comparison of the experimental data (symbols)
with the theoretical predictions of Eqs.  (\ref{eq:cyl_par_2}) and
(\ref{eq:cyl_perp_2}) represented by solid lines. The solid symbols represent
aggregates which consits of more than one string. All data were fitted with
Eqs.(\ref{eq:cyl_par_2}) and (\ref{eq:cyl_perp_2}) using the parameters given
in Sec.\ \ref{sec:experimental} except for the radius. This seemed to be
appropriate since the particles showed quite some dispersion. The fit
yielded $a = 17.4 \mu$m  $\pm 0.6 \mu$m, which is within the expected range.
Up to particle numbers of about 500 the fit is quite good, confirming the
validity of our simple model.  For higher values $n > 500$, the experimental
velocities are significantly larger than the theoretical prediction. Whereas
small aggregates (with 100 particles or less) consist of single vessels only,
these larger flotillas contain several vessels (up to four at $n \approx
1400$).\\


We next calculate the velocity of a large flotilla, assuming that its strings and fluid between them move with the same speed. Strictly speaking, this assumption is valid for close particle packing only, but for smaller area fraction of solids $\phi $ we should take into account hydrodynamic interactions between strings to evaluate their speed more accurately. In our simple theory a collective vessel translation is
modeled by the motion of a disc (which is the limiting case of an
oblate spheroid) under the force perpendicular to the axis of rotation.
 According to Ref.\ \onlinecite{happel1965},
Eq.  (5-11.25), for a thin disk of radius $R$ its velocity is given by
$ u \approx 3F_{nN}/32  R \mu_{\mbox{\tiny eff}}$,
provided $a \ll R$. In our case, the
disk radius can be approximated as $R\approx a(nN/\phi )^{1/2}$ where $N$ is
the number of vessels. For a
given $\phi$, and using $F_{nN} = nN|\mathbf{f}|$, we obtain
\begin{equation}
u= C \frac{3}{4} \pi \sqrt{n N \phi}
\label{eq:disk}
\end{equation}
with $C$ defined as above (Eq. (\ref{eq:C})). Hence the disk velocity
scales as $n^{1/2}$, i.e., it grows much faster with $n$ than in the
case of single strings. Even more importantly, the flotilla consists
of $N$ vessels, which further increases its speed compared to isolated
vessels.

We recall that the data in Fig.\ \ref{fig:results1} indicate that in the flotilla regime the
particle transport is anisotropic, and the speed of flotillas is highest in the longitudinal (i.e. parallel to
$\mathbf{B}_0$) direction. However, Eq.\
(\ref{eq:disk}), which oversimplifies hydrodynamic interactions between vessels, obviously predicts an isotropic behavior.
We also note that the shape of flotillas is often
elliptical, rather than circular (see \ Fig.\ \ref{fig:exp2}). Nevertheless, Eq.\
(\ref{eq:disk}) provides an explanation of a much higher speed of the
flotillas compared to isolated vessels. Indeed, the two green dashed curves included in Fig.\ \ref{fig:results1} plot
the predictions of Eq.\ (\ref{eq:disk}) for two values of the solid area
fraction, $\phi$, $\phi = 0.02$ (lower curve) and $\phi=0.06$ (upper curve), and we see that theoretical results agree well with the experimental data.
Attempts to estimate $\phi$ experimentally turned out difficult due to the
variable shapes of the flotillas.
Forcing a circular shape gave the extimate $\phi < 0.1$, however with an error
of the same size. Thus, the consideration above is quite approximate,
but it provides us with some guidance.

Finally we note that in all cases considered here (Eqs.\ (\ref{eq:cyl_par_2}),
(\ref{eq:cyl_perp_2}), Eq.\ (\ref{eq:disk})), the velocity scales
like $u \propto C \propto a^2$ with the particle size, due to the fact
that the magnetic force on the disk is proportional to its volume
while the drag is proportional to its radius. Specifically, the disc
velocity for total particle volume $V=4\pi a^{3}nN/3$ can be
written as
\begin{equation}
u=\frac{\rho MG(3\pi aV\phi )^{1/2}}{16\mu_{\mbox{\tiny eff}} }.
\label{eq:disk2}
\end{equation}
Since $u \propto a^{1/2}$, the collective velocity of vessels
moving in a disk/flotilla should be larger for bigger particles.

\section{Conclusion}


To summarize, we have shown that magnetic particles can be manipulated
and steered very efficiently by magnetic fields using our setup of combined
strong dipolar and weak quadrupolar magnetic fields, much more efficiently than
one would expect based on the Stokes friction of single particles. The reason
is that the dipolar field induces particle orientation and subsequent
aggregation, and that these aggregates then move as a whole.  The aggregates are hierarchically organized with
strings or spindles (vessels) being the lowest level entities, which are
further organized in flotillas.  The packing of particles in vessels already leads to their high translational speed
in the presence of a steering field. However, the formation of flotillas of many vessels further and dramatically increases the velocity response of particles
to the steering field.

We have provided a theoretical interpretation of our data, and found good
agreement between our model and experiment. Our theoretical model predicts that the velocity of vessels at fixed
number of constituent particles $n$ should scale as $\ln(n)$, whereas
the velocity of disks increases with $\sqrt{n}$. This explains why disks
are much faster at large $n$.

In the situation considered in the present work, the particles were
located at water-air interface, so that the shape of the flotillas
was two-dimensional (disks). Preliminary results in bulk solutions suggest that
a three-dimensional association of vessels is also possible (unpublished).  An
association into sphere-shaped flotillas should speed up the particles even
further.  In that case the velocity of the fully immersed in water sphere of radius $R=a(nN/\varphi
)^{1/3}$, where $\varphi $ is the volume fraction of solids, would scale as
$N^{2/3}$ following
\begin{equation}
u=\frac{F_{nN}}{6\pi R\mu }=\frac{2a^{2}\rho MG(nN)^{2/3}\varphi ^{1/3}}{%
9\mu }.  \label{13}
\end{equation}%
For given total particle volume $V$ and particle volume fraction
$\varphi$, the velocity then becomes independent of particle size:
\begin{equation}
u=\frac{2\rho MG(3V/4\pi )^{2/3}\varphi ^{1/3}}{9\mu }.  \label{15}
\end{equation}%
Thus superfast transport can be achieved for particles of any size
in this case.

The remaining open question is why the flotillas form at all.  In Monte
Carlo simulation of monodisperse, purely repulsive, dipolar particles, no such
association was observed (unpublished data).  As mentioned in the main text,
chain defects may provide a possible explanation.  Such defects could be
induced, e.g., by the polydispersity of particles.  Alternatively, the
flotillas may be stabilized by elastic interactions due to deformations of the
air-water interface. This interaction mechanism would however be restricted to
particles at interfaces. As we have shown, the association of magnetic
particles to flotillas highly enhances their response to external steering
fields, and they can thus be manipulated much more efficiently. In future work,
we thus plan to identify and investigate possible association mechanisms in
order to be able to control and exploit them.


\begin{acknowledgments}

This work was funded by the German Research Foundation (DFG), SFB 1066, within
projects A3, B5 and Q1. PB and OB wish to acknowledge additional funding by a
grant of the Johannes Gutenberg University Mainz (Inneruniversit\"are
Forschungsf\"orderung). ESA and OIV acknowledge support of the Russian Ministry of Education and Science.

\end{acknowledgments}

\bibliographystyle{apsrev4-1}
\bibliography{part}

\end{document}